\documentclass[a4paper]{jpconf}
\usepackage{graphicx}
\usepackage{amsmath}
\usepackage[
  colorlinks, 
  pdftitle=ERE14proceedings\ by\ C\ Schell\ \&\ O\ Rinne\ , 
  pdfsubject=based\ on\ a\ talk\ given\ by\ C\ Schell\ at\ ER14\ in\ Valencia, 
  pdfauthor=Christian\ Schell\ \&\ Oliver\ Rinne, 
  pdfkeywords=axisymmetry\ \&\ spectral\ \&\ fully\ constraint\ \&\ gauge, 
  pdfcreator=Kile\ \&\ nano, 
  pdfproducer=Makefile\ with\ latex\ \&\ bibtex\ \&\ dvips\ \&\ ps2pdf, 
  pdftoolbar=true]{hyperref} 
\hypersetup{citecolor=blue, linkcolor=blue, urlcolor=blue}

\begin{document}

\title{Spectral approach to axisymmetric evolution of Einstein's equations}

\author{Christian Schell and Oliver Rinne}

\address{Max Planck Institute for Gravitational Physics (Albert Einstein Institute),\break Am M\"uhlenberg 1, 14476 Golm, Germany and\break
Department of Mathematics and Computer Science, Freie Universit\"at Berlin,\break
Arnimallee 26, 14195 Berlin, Germany}

\ead{christian.schell@aei.mpg.de, oliver.rinne@aei.mpg.de}

\begin{abstract}
 We present a new formulation of Einstein's equations for an axisymmetric 
 spacetime with vanishing twist in vacuum. 
 We propose a fully constrained scheme and use spherical polar coordinates. 
 A general problem for this choice is the occurrence of coordinate 
 singularities on the axis of symmetry and at the origin. 
 Spherical harmonics are manifestly regular on the axis and hence take care of 
 that issue automatically. 
 In addition a spectral approach has computational advantages when the equations
 are implemented.
 Therefore we spectrally decompose all the variables in the appropriate 
 harmonics. 
 A central point in the formulation is the gauge choice.
 One of our results is that the commonly used maximal-isothermal gauge turns 
 out to be incompatible with tensor harmonic expansions, and we introduce a new
 gauge that is better suited.
 We also address the regularisation of the coordinate singularity at the origin.
\end{abstract}

\section{Introduction}
\label{introduction}

In this paper we consider the vacuum Einstein equations under the assumption of 
axisymmetry, i.e. there is an everywhere spacelike Killing vector field with 
closed orbits, $\partial_\varphi$ in spherical polar coordinates 
$(t,r,\vartheta,\varphi)$ adapted to the symmetry.
As an initial step we further assume that $\partial_\varphi$ is 
hypersurface-orthogonal (or twist-free), so there is no rotation.

The motivation for considering this situation comes from the obvious fact that 
due to the reduced dimensionality, it is much less computationally demanding than 
the case without symmetries. 
On the other hand we are not oversimplifying too much in the sense that our 
situation shares important features and properties of the full theory
such as the existence of gravitational waves.
In fact it was shown in~\cite{bicakpravdova98} that it 
is not possible to assume any further reasonable symmetry 
in the given situation when demanding gravitational radiation.
Due to the so-called Birkhoff theorem (for interesting historic remarks 
see~\cite{johansenravndal06}),
vacuum spherical symmetry is non-dynamical and thus not of interest to us.

We focus on isolated systems here, i.e.~we assume spacetime is asymptotically 
flat.
This provides us with appropriate fall-off conditions that can be used at an 
artificial boundary far away from the situation of interest.

There are many interesting applications we have in mind.
These include the collapse of gravitational waves, in particular critical 
phenomena.
Up to date there are, to the best of our knowledge, only two successful 
implementations finding critical phenomena for the vacuum axisymmetric Einstein 
equations~\cite{abrahamsevans93,sorkin11}.
These studies obtained rather different results for different initial data,
and further investigations are needed.
Another interesting question in mathematical relativity is the one of 
stability of solutions such as black holes.
Our setup will allow us to study the evolution of perturbations compatible
with our symmetry assumptions.

Our aim here is to derive a fully constrained formulation in spherical polar
coordinates which may be implemented by making use of a spectral approach for 
the angular part.
A well-known fact is that only six of the ten Einstein equations are of 
dynamical character, the other four are constraints.
An approach that is often applied to numerical investigations of Einstein's 
equations is to solve the constraints only on an initial hypersurface 
and then to evolve the system according to the remaining evolution equations.
This gives rise to the so-called free evolution.
Analytically the constraints remain preserved, which justifies the 
approach.
Numerically, on the other hand, there may exist constraint-violating modes, 
which cause instabilities.
Therefore our approach consists in enforcing the constraints at each timestep.
By construction there are no constraint-violating modes then.
Such formulations are called fully constrained, for previous publications see 
e.g.~\cite{bonazzolaetal04,corderocarrionetal08} for the case without symmetries
and~\cite{rinne08} for the case of axisymmetry.
A disadvantage of solving the constraints is that they are of elliptic nature 
and hence computationally much more involved.
We will introduce ideas how to save computational cost at other points in the 
formulation.

This brings us to the choice of a coordinate system.
At least for the implementation it is very common to introduce a fixed 
coordinate system.
In contrast to many previous formulations for similar situations which use 
cylindrical polar coordinates,
we choose spherical polar coordinates $(t,r,\vartheta,\varphi)$.
Besides the motivation from astrophysics, where many objects have an 
approximate spherical shape, the main reason is a mathematical one.
If spacetime has a topology $\mathcal{M}^2\times S^2$ then spherical harmonics
$Y_{\ell m}(\vartheta,\varphi)$ form an appropriate system for the spectral 
expansion on the sphere~$S^2$.
Another nice property is that fall-off conditions to be imposed at the outer 
boundary are usually given as an expansion in inverse powers of~$r$.
For the choice of spherical coordinates and spectral expansion in general relativity see for example~\cite{bonazzolaetal99,csizmadiaetal13,szilagyi14}.

In these proceedings we mainly focus on conceptual issues of the formulation, 
in particular the gauge choice. 
In section~\ref{spectralapproach} we introduce our spectral expansion.
It will be applied in section~\ref{formulation}, where we describe the 
derivation of the nonlinear equations, further their linearisation and 
regularisation.
In particular we focus on the gauge choice and investigate two possibilities in 
detail.
In the last chapter we briefly summarize and give an outlook on work in 
progress.

For us general relativity is given by the Einstein equations on a 
four-dimensional Lorentzian manifold $(\mathcal{M}^4, g)$ which is metric 
compatible, torsion-free and globally hyperbolic. 
Indices $i,j$ run from 1 to 3.

\section{Spectral approach}
\label{spectralapproach}

A general problem when using spherical polar coordinates is the occurrence of coordinate singularities 
at the origin ($r=0$) and the axis of symmetry ($\vartheta=0,\pi$).
As an example consider the flat Laplacian in spherical coordinates,
\begin{gather}
  \label{generallaplacian}
  \triangle = \partial_r^2+\frac{2}{r}\partial_r+\frac{1}{r^2}\left(\partial_\vartheta^2+\frac{\cos\vartheta}{\sin\vartheta}\partial_\vartheta\right).
\end{gather}
In fact many equations contain operators that have some similarity with the 
Laplacian, which is why we will take it as a model to illustrate our ideas. 

Since spherical harmonics are regular on the axis, they take care of that issue 
automatically.
Because of the assumed twist-free axisymmetry, all quantities are 
$\varphi$-independent.
This implies in particular that the spherical harmonics reduce to the $m=0$ 
harmonics
\begin{gather}
 \label{scalarharmonics}
 Y:=Y_\ell(\vartheta)=\sqrt\frac{2\ell+1}{4\pi}P_\ell(\cos\vartheta)
\end{gather}
with Legendre polynomials $P_\ell(\cos\vartheta)$. 
We shall omit the index $\ell$ if it is clear from the context.
However general relativity is a tensor theory.
We need to know the particular behaviour of each component of a scalar 
function, a vector and a symmetric two-tensor \cite{sarbachtiglio01,rinne09}.
Let $\hat g _{AB}$ be the round metric on the unit sphere and
$\hat \nabla$ its covariant derivative ($A,B=\vartheta,\varphi$).
The general even parity quantities are $Y_A:=\partial_A Y$ and 
$Y_{AB}:= [ \hat \nabla_A \hat \nabla_B Y ]^\mathrm{tf}$, the odd parity quantities play no role because of hypersurface-orthogonality.
In the case of twist-free axisymmetry the relevant basis harmonics are given by
\begin{subequations}
\label{scalarvectortensorharmonics}
 \begin{gather}
  Y = {}_0Y\label{scalarharmonics},\\
  Y_\vartheta=-\frac{1}{2}\sqrt{\ell(\ell+1)}\left({}_1Y-{}_{-1}Y\right)\label{vectorharmonics},\\
  Y_{\vartheta\vartheta}
  = -\left[\frac{\cos\vartheta}{\sin\vartheta}Y_\vartheta+\frac{\ell(\ell+1)}{2}Y\right]
  =\frac{1}{4}\sqrt{(\ell-1)\ell(\ell+1)(\ell+2)}\left({}_2Y+ {}_{-2}Y\right),
  \label{tensorharmonics}
 \end{gather}
\end{subequations}
where we have used the properties of the Legendre functions to
eliminate the second $\vartheta$-derivatives in~\eqref{tensorharmonics}.
For completeness we have also given the expressions in terms of the
spin-weighted harmonics ${}_{s}Y$ \cite{rinne09}.
In the following we will refer to $Y$, $Y_\vartheta$ and $Y_{\vartheta\vartheta}$ as 
the scalar, vector and tensor harmonics, respectively.
Fields expanding in those harmonics will be called scalar, vector and 
tensor quantities.

We explicitly give the expansion of some components of the spatial metric 
$\gamma_{ij}$ needed in the following:
\begin{subequations}
\label{gammaexpansion}
 \begin{gather}
  \gamma_{rr} = H Y,\\
  \gamma_{\vartheta\vartheta}=r^2 \left(K - \frac{\ell(\ell+1)}{2} G\right) Y - r^2 \frac{\cos\vartheta}{\sin\vartheta} G Y_\vartheta
   =r^2\left(KY+GY_{\vartheta\vartheta}\right),\\
  \gamma_{\varphi\varphi}=r^2 \sin^2\vartheta \left(K + \frac{\ell(\ell+1)}{2} G\right) Y + r^2\cos\vartheta\sin\vartheta G Y_\vartheta
   =r^2\sin^2\vartheta\left(KY-GY_{\vartheta\vartheta}\right),
 \end{gather} 
\end{subequations}
where $H, K$ and $G$ are functions of $t$ and $r$ only and a sum over 
$\ell$ is implied.

\section{Formulation}
\label{formulation}

Our starting point is the usual 3+1 decomposition of general relativity $(\mathcal{M}^4,g)\mapsto(\Sigma^3,\gamma,K)$ 
\cite{york79,gourgoulhon12}.
Here $\Sigma^3$ is a level set of three dimensional spacelike hypersurfaces and 
$\gamma$ and $K$ are their first and second fundamental forms.
The evolution takes place along the timelike vector field~$t$ (recall that 
$(\mathcal{M}^4,g)$ is globally hyperbolic by assumption).
The line element is given by 
\begin{gather}
 \textrm{d}s^2=-\alpha^2\textrm{d}t^2+\gamma_{ij}\left(\textrm{d}x^i+\beta^i\textrm{d}t\right)\left(\textrm{d}x^j+\beta^j\textrm{d}t\right),
\end{gather}
where $\alpha$ is the lapse function and $\beta^i$ the components of the shift 
vector.
We obtain six evolution equations each for $\gamma_{ij}$ and $K_{ij}$, 
a Hamiltonian constraint and three momentum constraints.

Now, in this setting, twist-free axisymmetry means that all variables are 
$\varphi$-independent, $\gamma_{r\varphi}=\gamma_{\vartheta\varphi}=0$ and 
$\beta^\varphi=0$.
These identities are preserved under time evolution.
It follows that also $K_{r\varphi}$ and $K_{\vartheta\varphi}$ have to vanish and 
the $\varphi$-momentum constraint is identically satisfied.

The diffeomorphism invariance of general relativity is encoded in the lapse and 
shift.
To fix the gauge we have to choose a slicing condition and two spatial gauge 
conditions.

\subsection{Choice of a gauge}
\label{gaugechoice}

An evident choice is the so-called maximal-isothermal gauge, 
see~\cite{dain11} for a review.
This is a combination of maximal slicing, $\textrm{tr}K=0=\partial_t\left(\textrm{tr}K\right)$, and the 
quasi-isotropic condition.
The latter one consists of the diagonal gauge, 
\begin{gather}
 \label{diagonalgauge}
 \gamma_{r\vartheta}=0=\partial_t\gamma_{r\vartheta},
\end{gather}
and a condition that separates the remaining $\varphi$-part of the spatial 
metric by relating the other components as
\begin{gather}
 \label{oldgauge}
 \gamma_{\vartheta\vartheta}=r^2\gamma_{rr},
\end{gather}
which is also preserved in time, 
$\partial_t\left(\gamma_{\vartheta\vartheta}-r^2\gamma_{rr}\right)=0$.
This is a widely used gauge in axisymmetric simulations 
\cite{abrahamsevans93,garfinkleduncan01,choptuiketal03,rinne08}
and also analytically well studied~\cite{dain11}.
As we will see later in section~\ref{linearization} after the linearisation 
and expansion in spherical harmonics, the maximal-isothermal gauge
is unfortunately \emph{not} an appropriate gauge for our purposes.
In order to find a well-suited condition we decide to keep maximal slicing 
and the diagonal gauge as before but to come up with a new condition
for the diagonal components of metric, namely
\begin{gather}
 \label{newgauge}
 \gamma_{\vartheta\vartheta}=r^4\sin^2\vartheta\ \gamma^{\varphi\varphi}(\gamma_{rr})^2.
\end{gather}
This gauge should also be preserved in time, 
$\partial_t\left(\gamma_{\vartheta\vartheta}-r^4\sin^2\vartheta\ \gamma^{\varphi\varphi}(\gamma_{rr})^2\right)=0$.
Note that the nonlinear condition~\eqref{newgauge} relates all the remaining 
components of the spatial metric.
We will show in section~\ref{linearization} that this is indeed an appropriate 
gauge for our purposes.
The three preservation equations in $t$ for the gauge choices give us further 
elliptic equations to be solved at each time step in the fully constrained system.

Having the new gauge condition at hand we can now follow the usual procedure in the $3+1$~formulation of general relativity 
to derive the nonlinear constraints and evolution equations.
For obvious reasons one should choose variables in such a way that their 
linearisation has a convenient expansion in scalar, vector and tensor harmonics 
in the sense explained at the end of section~\ref{spectralapproach}.
We note that it is sometimes useful to add appropriate multiples 
of the constraints (which vanish for a solution to Einstein's equations)
to some of the equations such that the linearisation of the equations also 
expand in a definite way.

We have eight variables, namely the lapse function, two components of the shift 
vector, two components of the spatial metric and
three components of the extrinsic curvature.
On the other hand we have six elliptic equations: the Hamiltonian and two momentum 
constraints, the preservation of maximal slicing,
the diagonal gauge condition~\eqref{diagonalgauge} and the newly introduced 
gauge condition~\eqref{newgauge}.
Since we are looking for a fully constrained formulation, we will explicitly 
solve all of these elliptic equations in our scheme.
Furthermore we have five evolution equations, two for the remaining components of the spatial metric~$\gamma_{ij}$ and 
three for the extrinsic curvature~$K_{ij}$.
Thus the system is overdetermined.
In the following we will concentrate on those two evolution equations for the 
components of $\gamma_{ij}$ and $K_{ij}$ that expand in the linearisation in 
tensor harmonics.
Besides being reasonable to choose two canonically conjugated variables as
evolved fields, we expect the tensor quantities to carry the gravitational 
wave degrees of freedom (at least in linearised theory).
The other evolution equations may be used for consistency checks 
but will not be considered in the remainder of this paper.

\subsection{Linearisation}
\label{linearization}

Having derived the equations, we next linearise them about a flat background
spacetime.
This means we expand all quantities in the form 
$f=f|_\textrm{flat}+\epsilon\tilde{f}$ and just keep terms of linear order in 
$\epsilon$, ignoring higher-order terms.
We obtain two evolution equations and six constraints, all dependent on 
$(t,r,\vartheta$).
As expected we are faced with singularities both on the axis and at the origin.
One should think of the Laplacian in~\eqref{generallaplacian} as a model
operator.

On the linearised level we expand all variables in the corresponding spherical 
harmonics as given at the end of section~\ref{spectralapproach}.
E.g.~for a variable $\tilde{f}$ that expands in scalar harmonics, we have
\begin{gather}
 \label{sphericalexpansion}
 \tilde{f}(t,r,\vartheta)=\sum_\ell\hat{f}_\ell(t,r)Y_\ell(\vartheta).
\end{gather}
Applying the expansion of the metric coefficients \eqref{gammaexpansion} 
to the quasi-isotropic condition~\eqref{oldgauge}, one finds
\begin{gather}
 r^2\left(H-K+\frac{\ell(\ell+1)}{2}G\right)Y + r^2\frac{\cos\vartheta}
 {\sin\vartheta} GY_\vartheta = 0,
\end{gather}
which implies $G=0$ and hence $H=K$.
Thus only one degree of freedom for the spatial metric remains.
Therefore the only situation that is compatible with this choice is the one of 
spherical symmetry.

On the other hand, linearising~\eqref{newgauge} about flat space leads to the 
condition
\begin{gather}
 \gamma_{\vartheta\vartheta}=2r^2\gamma_{rr}
 -\frac{\gamma_{\varphi\varphi}}{\sin^2\vartheta}
\end{gather}
and hence, again by using~\eqref{gammaexpansion},
\begin{gather}
 2r^2 (K - H) Y = 0.
\end{gather}
Therefore $H=K$ and $G$ arbitrary are two remaining degrees of freedom, 
which shows that these conditions are indeed well suited.

The $\vartheta$-dependence is completely absorbed in the spherical harmonics.
Thus the interesting part is now contained in $\hat{f}_\ell$ 
in~\eqref{sphericalexpansion}, which depends on $(t,r)$ only.
On the linear level one expects a decoupling of all the different $\ell$-modes.
This is indeed the case provided one considers the ``correct'' nonlinear 
equations as explained in the previous subsection~\ref{gaugechoice},
adding appropriate multiples of the constraints.
Therefore we obtain, for each $\ell$-mode, a $(1+1)$-dimensional system of 
equations in $(t,r)$.
The equations are still formally singular at the origin $r=0$.

\subsection{Regularisation}

In order to regularise the just obtained equations at the origin $r=0$, 
we follow a procedure proposed 
e.g.~in~\cite{grandclementnovak09,gundlachetal13}.
Let us again take the Laplace operator~\eqref{generallaplacian} as an example.
After expansion in (scalar) spherical harmonics, the Laplace equation reads 
\begin{gather}
 \label{expandedlaplacian}
 \partial_r^2\hat{f}_\ell+\frac{2}{r}\partial_r\hat{f}_\ell-\frac{\ell(\ell+1)}{r^2}\hat{f}_\ell=0.
\end{gather}
This equation can be solved explicitly.
Its solution is a superposition of a singular part proportional to 
$r^{-(\ell+1)}$ and a regular part proportional to~$r^\ell$.
Since we only consider solutions that are smooth at $r=0$,
we require the integration constant of the part proportional to 
$r^{-(\ell+1)}$ to vanish.
If we isolate the leading-order behaviour by setting 
$\hat{f}_\ell(t,r)=:r^\ell\bar{f}_\ell(t,r)$,
we can continue to work with the barred quantities, which expand, 
close to the origin $r=0$, in even power series in $r$.
Therefore \eqref{expandedlaplacian} is now manifestly regular,
\begin{gather}
 \partial_r^2\bar{f}_\ell+\frac{2}{r}(\ell+1)\partial_r\bar{f}_\ell=0.
\end{gather}
For our set of equations we have to pull out factors of either $r^\ell$, $r^{\ell+1}$ 
or $r^{\ell+2}$ to obtain similar results,
but indeed, the system can be completely regularised in this way.

A further nice property is that, with a little bit of rearranging, one obtains 
a hierarchy of equations.
For the implementation we evolve, from one time step to the next, 
the two tensor variables by using the evolution equations.
Here we remark that, when taking the second time derivative of the tensor 
component of~$\gamma_{ij}$ and using the evolution equation for the corresponding 
component of $K_{ij}$,
the principal part of the equations is just the ordinary wave equation.
Then, on the new time level, we solve successively the constraints to obtain 
the remaining variables on that time slice before stepping to the next slice.
Given the tensor part of~$\gamma_{ij}$, the Hamiltonian constraint gives us the 
scalar component of~$\gamma_{ij}$, 
while the preservation of maximal slicing determines the lapse function.
Given the tensor contribution to~$K_{ij}$,
the two momentum constraints allow us to determine the remaining 
(scalar and vector) components of the extrinsic curvature.
Finally the preservation of the spatial gauge conditions determines the 
components of the shift vector.

\section{Conclusion}

We have formulated Einstein's equations for an isolated system in twist-free 
axisymmetry.
Key features of our formulation are that the scheme is fully constrained and 
uses a spherical polar coordinate system.
In its linearisation about a flat background, a spectral expansion in spherical 
harmonics may be used and the resulting equations are fully regularisable at the origin.
A~central result presented in this paper is the fact that the well-understood 
and frequently applied maximal-isothermal gauge is not compatible with 
tensor spherical harmonic expansions.
Instead we proposed another gauge condition which is well-suited.
Currently we are in the process of coding the linearised system and 
finding a proper way to include the non-linearities.
Possible directions for further studies include a deeper analysis of the 
properties of the system, the inclusion of a non-vanishing twist to allow for 
rotating spacetimes and, ultimately,
the application of the code to physically and mathematically interesting 
situations such as those mentioned in section~\ref{introduction}.

\ack{This research is supported by grant RI 2246/2 of the German Research 
Foundation (DFG) and a Heisenberg Fellowship to OR.
CS also acknowledges support from the International Max Planck Research 
School (IMPRS) for Geometric Analysis, Gravitation and String Theory.}

\section*{References}

\bibliography{paper} 

\end{document}